\documentclass[aps,pra,twocolumn,superscriptaddress]{revtex4-1}
\usepackage{amsmath,amsfonts,amssymb}
\usepackage{graphicx,float,calc}
\usepackage{color,bm}
\usepackage{ulem}
\usepackage{braket}\hyphenpenalty=5000
\usepackage[colorlinks,urlcolor=blue,citecolor=blue,linkcolor=blue]{hyperref}

\begin{document}
	
\title{Antichiral and trap-skin dynamics in a nonreciprocal bosonic two-leg ladder with artificial magnetic flux}

\author{Rui-Jie Chen}
\affiliation{Key Laboratory of Atomic and Subatomic Structure and Quantum Control (Ministry of Education), Guangdong Basic Research Center of Excellence for Structure and Fundamental Interactions of Matter, South China Normal University, Guangzhou 510006, China}
\affiliation{Guangdong Provincial Key Laboratory of Quantum Engineering and Quantum Materials,School of Physics, South China Normal University, Guangzhou 510006, China}
\author{Guo-Qing Zhang}
\email{zhangptnoone@zjhu.edu.cn}
\affiliation{Research Center for Quantum Physics, Huzhou University, Huzhou 313000, People's Republic of China}
\author{Zhi Li}
\affiliation{Key Laboratory of Atomic and Subatomic Structure and Quantum Control (Ministry of Education), Guangdong Basic Research Center of Excellence for Structure and Fundamental Interactions of Matter, South China Normal University, Guangzhou 510006, China}
\affiliation{Guangdong Provincial Key Laboratory of Quantum Engineering and Quantum Materials, School of Physics, South China Normal University, Guangzhou 510006, China}
\author{Dan-Wei Zhang}
\email{danweizhang@m.scnu.edu.cn}
\affiliation{Key Laboratory of Atomic and Subatomic Structure and Quantum Control (Ministry of Education), Guangdong Basic Research Center of Excellence for Structure and Fundamental Interactions of Matter, South China Normal University, Guangzhou 510006, China}
\affiliation{Guangdong Provincial Key Laboratory of Quantum Engineering and Quantum Materials, School of Physics, South China Normal University, Guangzhou 510006, China}

\date{\today}	

\begin{abstract}
Non-Hermiticity and synthetic gauge fields play two fundamental roles in engineering exotic phases and dynamics in artificial quantum systems. Here we explore the mean-field dynamics of interacting bosons in a two-leg ladder with synthetic magnetic flux and nonreciprocal hopping under the open boundary condition. In the Hermitian limit, we showcase the breakdown of the flux-driven chiral dynamics due to the nonlinear self-trapping effect. We further find that the nonreciprocity can drive the transition between chiral dynamics and antichiral dynamics. The antichiral motion is manifested as the non-Hermitian skin dynamics along the same direction on two legs that are not suppressed by the magnetic flux, while the chiral-antichiral transition is flux-tunable. We also reveal the trap-skin dynamics with the coexistence of the self-tapping and skin dynamics in the ladder. Dynamical phase diagrams with respect to the chiral-antichiral dynamics, skin dynamics, self-trapping dynamics, and trap-skin dynamics are presented. Our results shed light on intriguing dynamical phenomena under the interplay among non-Hermiticity, nonlinearity, and artificial gauge fields.
\end{abstract}	

\maketitle

\section{\label{sec1}Introduction}

Hermiticity in quantum mechanics ensures real energy spectra and probability conservation in isolated quantum systems. In the past decade, non-Hermitian classical and quantum systems have attracted increasing attention~\cite{Hatano1996,Hatano1997,Bender1998,Ueda2020,Bergholtz2021}. There are two typical kinds of non-Hermitian systems, where the non-Hermiticity comes from the gain-and-loss and nonreciprocality, respectively. Extensive intriguing properties unique to these non-Hermitian systems have been discovered, such as exceptional points ~\cite{Bergholtz2021,Shen2018,Xue2021,Tzeng2021,Ding2022,Sticlet2022,Sticlet2023,li2023exceptional}, new topological phases and invariants~\cite{Yao2018Aug,Yao2018Sep,Gong2018,Song2019,Jin2019,Xue2020,wang2021complex-energy,wang2021generating,DW2020TAI,Tang2020,Xue2022TAI}, the non-Hermitian skin effect (NHSE)~\cite{Yao2018Aug,Jiang2019,li2020critical,DW2020Skinsuperfluid,zhang2021observation,Zhang2021,YYi2020,LLi2020,zhang2022universal,Longhi2022,Longhi2022selfhealing,Xiujuan2022,Xue2022edgeburst,Li2022Direction,Zhang2023,Kawabata2023,lin2023topological,XWang2023,Ezawa2022,SZLi2024NHSE} and mobility ring~\cite{Wang2024NonHermitianBS,li2024ring}.
The so-called NHSE is manifested as skin modes (eigenstates) localized near the boundaries and profoundly modifies the bulk-boundary correspondence in non-Hermitian systems \cite{Yao2018Aug}.
In Ref.~\cite{Ezawa2022}, the skin modes were shown to be essentially modified by an additional nonlinear term in the Hatano-Nelson model \cite{Hatano1996,Hatano1997}. Notably, the interplay between non-Hermiticity and nonlinearity in non-Hermitian systems remains largely unexplored~\cite{Konotop2016,Xia2021,Graefe2008,DW2021nonlinear,Zhao2020,Lang2021,Hang2021,Pyrialakos2022,Rahmani2024}.

On the other hand, many efforts have been made to engineer artificial gauge fields in ultracold atoms~\cite{Dalibard2011,Goldman2014,Goldman2016,DW2018,Cooper2019,Lin2009,Aidelsburger2013,Miyake2013,Yan2022} and photonic systems \cite{lu2014topological,Khanikaev2017,Ozawa2019,dutt2020single}. For instance, one of the simplest systems to study dynamical and topological phenomena is two-leg ladders with effective magnetic fields~\cite{Dariol2014}. Various tunable optical ladders with synthetic magnetic fluxes have been experimentally realized for ultracold atoms ~\cite{Atala2014,Tai2017,Bryce2017Direct,Suotang2022Atom,Suotang2022,Suotang2023,Yan2024EngineeringTC}. It has been demonstrated that the synthetic magnetic flux can induce chiral Meissner currents and vortex phases in the bosonic ladders in the absence~\cite{Atala2014} or presence of interatomic interactions \cite{Tai2017,Piraud2015,Zheng2017,Li2020,Zhang2022nonlinear,Giri2023}. Meanwhile, tunable nonreciprocal hopping for ultracold bosons has been realized through a dissipative Aharonov-Bohm chain with the synthetic magnetic flux \cite{Yan2020Tunable}, where the dynamic signatures of the NHSE have recently been observed \cite{Liang2022}. For noninteracting atoms in a two-leg ladder lattice with the on-site gain-and-loss, the flux-dependent NHSE and topological transitions were theoretically revealed \cite{Wu2022}. As a signature of the NHSE, the antichiral edge currents of noninteracting photons in a synthetic two-leg ladder with the effective magnetic flux and gain-and-loss were experimentally observed \cite{Ye2023ObservationON}. For two-dimensional non-Hermitian lattices, the (second-order) NHSE can be significantly suppressed (enhanced) by the magnetic fields \cite{Lu2021,KShao2022,Li2023enhancement}. A key problem in this direction is the effects of the interplay among non-Hermiticity, synthetic magnetic fields and interactions.

In this work, we numerically investigate the mean-field dynamics of interacting bosons in a two-leg ladder with open boundaries, in the presence of synthetic magnetic flux and nonreciprocal hopping. In the Hermitian limit, we show that the flux-driven chiral dynamics can be broken down by the self-trapping effect as the strength of the nonlinear interaction is increased, with a crossover region for the coexistence of chiral and trapping dynamics. We also find that the nonreciprocity can lead to the transition between the chiral dynamics and antichiral dynamics, which is tunable by the magnetic flux. In our ladder with open boundaries, the antichiral motion is manifested as the non-Hermitian skin dynamics along the same direction on two legs, which is not suppressed by the magnetic flux. Moreover, we reveal the trap-skin dynamics with coexistence of the self-tapping and skin dynamics. By numerically calculating various physical quantities, we obtain dynamical phase diagrams with respect to the chiral-antichiral dynamics, skin dynamics, self-trapping dynamics, and the trap-skin dynamics.

The rest of this paper is organized as follows. In Sec.~\ref{sec2}, we introduce the two-leg model with the nonreciprocal hopping and magnetic flux. Section \ref{sec3} is devoted to present our main results for various dynamical properties and related phase diagrams. Finally, a brief discussion on the finite-size effect and a short conclusion are given in Sec.~\ref{sec4}.

\section{\label{sec2}Model}
\begin{figure}[t!]
	\centering
	\includegraphics[width=0.45\textwidth]{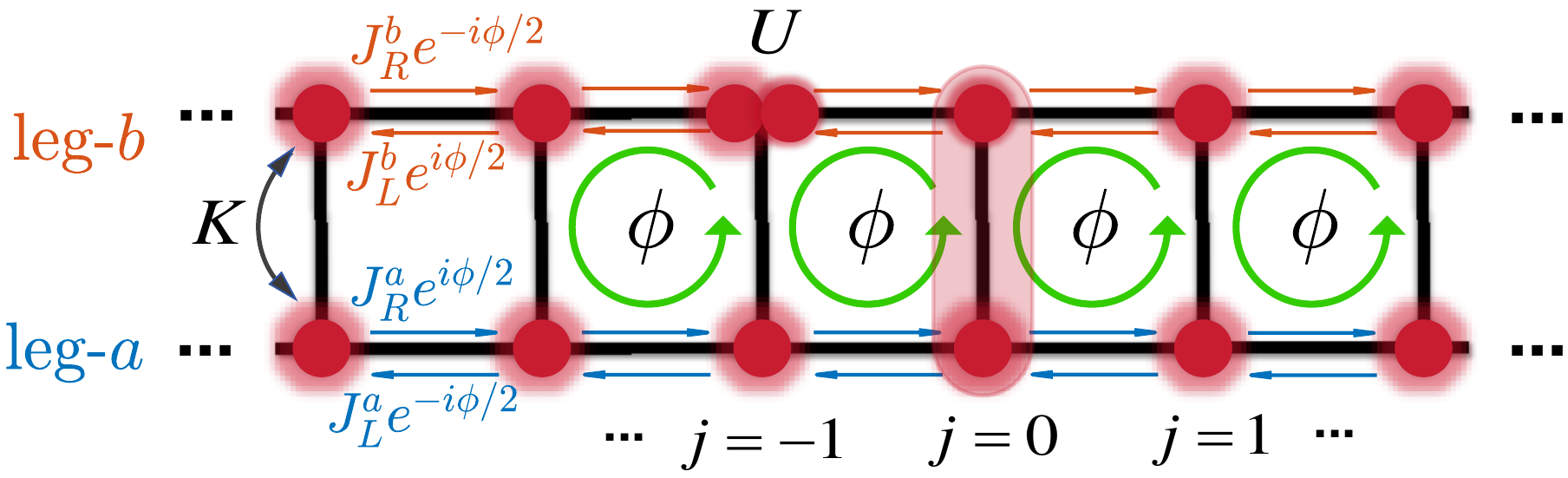}
	\caption{(Color online) Schematic of the two-leg ladder lattice of interacting bosons with the nonreciprocal hopping and an artificial magnetic flux denoted by $\phi$. Here $K$ denote the inter-leg hopping strength, $U$ is the on-site interaction strength, and $J_{L,R}^{a,b}$ represent the nonreciprocal hopping amplitudes. The translucent red shade denotes the initial wave packet described by Eq.~(\ref{initial state}).}
	\label{fig1}
\end{figure}

We begin by considering interacting bosons in a two-leg ladder lattice with nonreciprocal hopping and an artificial magnetic flux, as shown in Fig.~\ref{fig1}(a). The tight-binding Hamiltonian of the system reads
\begin{equation}\label{Ham}
	\begin{split}
		\hat{H}=&\sum_{j}^{}(J_R^a e^{i\phi/2}\hat{a}_{j+1}^{\dagger}\hat{a}_j + J_L^a e^{-i\phi/2}\hat{a}_{j}^{\dagger}\hat{a}_{j+1}\\
		        &+J_R^b e^{-i\phi/2}\hat{b}_{j+1}^{\dagger}\hat{b}_j + J_L^b e^{i\phi/2}\hat{b}_{j}^{\dagger}\hat{b}_{j+1})\\
		        &+K\sum_{j}^{}(\hat{a}_j^{\dagger}\hat{b}_j + h.c.) + \frac{U}{2}\sum_{j,\mu}^{}\hat{\mu}_j^{\dagger}\hat{\mu}_j^{\dagger}\hat{\mu}_j\hat{\mu}_j,
	\end{split}
\end{equation}
where $\hat{\mu}_j$ ($\hat{\mu}_j^{\dagger}$) with $\mu=\{a,b\}$ denotes the annihilation (creation) operator for a bosonic atom at the site $j$ of the leg-$\mu$. Here, $J_R^{\mu}=Je^{-g_{\mu}}$ and $J_L^{\mu}=Je^{g_{\mu}}$ are nonreciprocal right- and left-hopping amplitudes with $g_{\mu}$ as the nonreciprocal strength; $U$ is the on-site interaction strength, and $K$ denotes the tunneling amplitude between two legs; The artificial magnetic field piercing through the ladder rises from exponents $e^{\pm i\phi/2}$. Hereafter we set $J=1$ and $\hbar/J$ as the energy and time units, respectively. In our numerical simulation, we focus on the ladder of the length $L=21$ with the site indices $j\in[-\frac{L+1}{2},\cdots,0,\cdots,\frac{L+1}{2}]$. The length is close to that of the atomic ladders in practical experiments ~\cite{Atala2014,Tai2017,Bryce2017Direct,Suotang2022Atom,Suotang2022,Suotang2023,Yan2024EngineeringTC}.
The main dynamical properties in this ladder preserve for larger ladder lengths and the finite-size effect is discussed in Sec.~\ref{sec4}. In the noninteracting limit with $U=0$, the Hamiltonian can be written in the momentum $k$ space by Fourier transformation under the periodic boundary condition (PBC), corresponding to the following Bloch Hamiltonian
\begin{equation}\label{bloch H}
	H(k)=\left[\begin{array}{ccc}
		J_R^a e^{i\varphi_+} + J_L^a e^{-i\varphi_+} & K \\
	    K & J_R^b e^{-i\varphi_-} + J_L^b e^{i\varphi_-}
	     \end{array}\right],
\end{equation}
where $\varphi_{\pm}(k)=\phi/2\pm k$. %The Bloch Hamiltonian is more convenient than the real space one when calculating topological properties.

%%%%%%%%%%%%%%%%%%%%	
%%%%%%%%%%%%%%%%%%%%

We study the dynamics of the system under the mean-field approximation. In this case, the annihilation and creation operators in the Hamiltonian in Eq. (\ref{Ham}) can be approximately replaced by their expectation values with complex numbers: $\hat{\mu}_j\approx\langle\hat{\mu}_j\rangle\equiv\psi_j^{\mu}$ and  $\hat{\mu}_j^{\dagger}\approx\langle\hat{\mu}_j^{\dagger}\rangle \equiv\psi_j^{\mu\dagger}$.
The equations of motion can be obtained straightforward via the replacement of the bosonic operators with the complex numbers~\cite{POLKOVNIKOV2010}, which are given by the following nonlinear Schr\"odinger equations:
\begin{equation}\label{nonlinear equation}
	\begin{split}
		i\frac{\partial}{\partial t}\left[\begin{array}{ccc}\psi_j^a\\ \psi_j^b \end{array}\right]=&
		\left[\begin{array}{ccc}J_R^a e^{i\phi/2}\psi_{j-1}^a\\ J_R^b e^{-i\phi/2}\psi_{j-1}^b \end{array}\right]+
		\left[\begin{array}{ccc}J_L^a e^{-i\phi/2}\psi_{j+1}^a\\ J_L^b e^{i\phi/2}\psi_{j+1}^b \end{array}\right]\\&+K\left[\begin{array}{ccc}\psi_j^b\\ \psi_j^a \end{array}\right]+U\left[\begin{array}{ccc}\vert\psi_j^a\vert^2\psi_j^a\\ \vert\psi_j^b\vert^2\psi_j^b \end{array}\right].
	\end{split}
\end{equation}
Now $U$ denotes the nonlinear interacting strength. For $U=g_{a,b}=0$, our model reduces to the Hermitian two-leg ladder of noninteracting bosons under the artificial magnetic flux \cite{Dariol2014}, where the single-particle chiral motion has been experimentally observed. The chiral motion is due to the correlation between the sign of the group velocity and the leg towards which the particle is biased.

In the following, we numerically study the wave-packet dynamics in the ladder with non-Hermitian hoppings and synthetic magnetic flux in the presence (absence) of the nonlinear interaction. To study the non-Hermitian and nonlinear effects on the chiral dynamics, we focus on the initial state that is prepared at the two legs of the central rung ($j=0$) of the ladder and reads
\begin{equation}\label{initial state}
 \left[\begin{array}{ccc}\psi_j^a(t=0)\\ \psi_j^b(t=0)\end{array}\right]
 =\frac{1}{\sqrt{2}}\left[\begin{array}{ccc}\delta_{j,0}\\\delta_{j,0}\end{array}\right].
\end{equation}
Under this initial condition, we numerically integrate Eq. (\ref{nonlinear equation}) and obtain the time evolution of the wave function, which is renormalized in each time step. When studying the chiral dynamics, we choose the evolution time within which the wave packet does not touch the boundary and thus consider the open boundary condition (OBC). In the investigation of NHSE, we consider a long-time evolution under the OBC.

\section{\label{sec3}Results}
\subsection{Chiral dynamics versus self-trapping dynamics}

\begin{figure}[t!]
	\centering
	\includegraphics[width=0.46\textwidth]{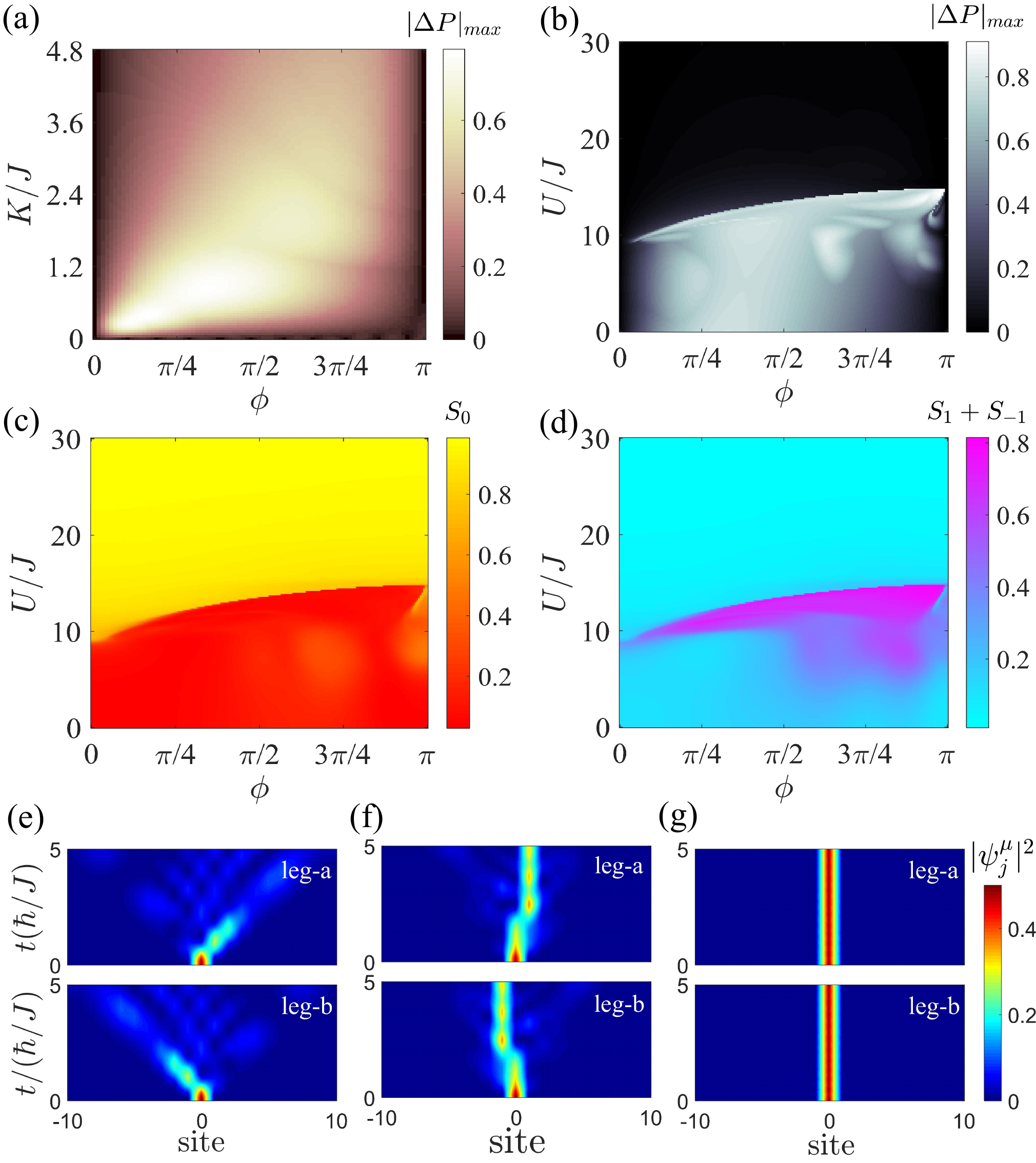}
	\caption{(Color online) (a) $|\Delta P|_{\mathrm{max}}$ as a function of $\phi$ and $K$ for  $U=0$. (b) $|\Delta P|_{\mathrm{max}}$, (c) $S_0$, and (d) $S_1+S_{-1}$ as functions of $\phi$ and $U$. Time evolution of the density distribution for (e) $U=0$, (f) $U=12$, and (g) $U=30$. Other parameters are $K=0.8$ in (b)-(g), $\phi=3\pi/8$ in (e)-(g), and $g_a=g_b=0$ in (a-g).}
	\label{fig2}
\end{figure}

In this section, we consider the nonlinear interaction effect on the chiral motion in the Hermitian limit of $g_a=g_b=0$. Since the chiral motion corresponds to an atomic current bias between two legs~\cite{Wu2022,Giri2023}, one can define the shearing
\begin{equation}\label{Delta P}
	\Delta P(t) = P_a(t) - P_b(t)
\end{equation}
to characterize the chiral dynamics, where $P_{\mu}$ is given by
\begin{equation}\label{P sigma}
	P_{\mu}(t) = \sum_{j>0}|\psi_j^{\mu}(t)|^2 - \sum_{j<0}|\psi_j^{\mu}(t)|^2.
\end{equation}
To avoid the boundary effect on the dynamics, we set the maximum evolution time $T=5$ (in the unit of $\hbar/J$) for the system of size $L=21$ and numerically compute the maximum shearing
\begin{equation}\label{max Delta P}
	|\Delta P|_{\mathrm{max}} = \mathop{\max}_{t\in[0,T]} |\Delta P(t)|
\end{equation}
during the evolution. The finite value of the maximum shearing implies the presence of the chiral dynamics. Figure \ref{fig2}(a) shows the numerical result of $|\Delta P|_{\mathrm{max}}$ as a function of $\phi$ and $K$ for $U=0$, which indicates that the chiral dynamics with $|\Delta P|_{\mathrm{max}}\neq0$ is exhibited if $\phi \neq 0,\pi$ in the noninteracting limit.

We vary the interacting strength $U$ for fixed $K=0.8$ to study the nonlinear effect. The result of $|\Delta P|_{\mathrm{max}}$ in the $\phi$-$U$ parameter plane is shown in Fig.~\ref{fig2}(b). One can find two parameter regions corresponding to the presence and absence of the chiral dynamics for small and large $U$, respectively. This indicates the breakdown of the chiral dynamics under strong nonlinear interactions \cite{Ezawa2022}. Notably, there is a disorder-free localization phenomenon in nonlinear systems, namely the self-trapping effect. To characterize this effect, we use the following quantity~\cite{Ezawa2022}
\begin{equation}\label{S}
	S_j=\frac{2}{T}\int_{T/2}^{T} (|\psi_j^a(t)|^2 + |\psi_j^b(t)|^2),
\end{equation}
which denotes the site-dependent density averaged over the time span from $T/2$ to $T$.
We choose $S_0$ to describe the self-trapping dynamics since the particle is initially placed at the $j=0$ site. In Fig.~\ref{fig2}(c), $S_0$ in the $\phi$-$U$ plane shows two parameter regions for the chiral and self-trapping dynamics, respectively, similar to those in Fig.~\ref{fig2}(b). In the self-trapping region, $S_0$ approaches to one and $|\Delta P|_{\mathrm{max}}$ tends to be vanishing due to the localization property. The self-trapping effect can suppress the population propagation along the radial direction and thus destroys the chiral motion. We further compute the population near the localization center as $S_1+S_{-1}$ in the $\phi$-$U$ plane in Fig.~\ref{fig2}(d). One can find a parameter region with significant values of $S_1+S_{-1}$, which corresponds to the crossover region with the coexistence of chiral and self-trapping dynamics. To be more clear, we show the time evolution of density distributions for various interacting strengths $U$ in Figs.~\ref{fig2}(e-g). For $U=0$ in Fig.~\ref{fig2}(e), the biased particle propagations on leg-$a$ and leg-$b$ toward the right and left sides of the ladder, respectively, show the chiral dynamics. Figure~\ref{fig2}(f) shows the trap-chiral dynamics for $U=12$: the biased particle propagations on different legs toward opposite sides but are trapped around the initial site. For the self-trapping dynamics with $U=30$ in Fig.~\ref{fig2}(g), the populations on both leg-$a$ and leg-$b$ are entirely trapped.

%%%%%%%%%%%%%%%%%%%%	
%%%%%%%%%%%%%%%%%%%%

\subsection{Non-reciprocity-induced antichiral dynamics}

We turn to investigate the dynamics in the presence of nonreciprocal hopping under the OBC. It has been shown that the on-site gain-and-loss in a noninteracting ladder can lead to the antichiral dynamics~\cite{Ye2023ObservationON,Wu2022}, as a result of the interplay between the flux-induced asymmetric transport and the inter-leg leaking. Under antichiral dynamics, atoms in the two legs propagate in the same direction, similar to the antichiral edge modes in two-dimensional Hermitian systems~\cite{Colomes2018, Mandal2019}. Notably, the antichiral dynamics can be induced by the NHSE in our nonreciprocal ladder under the OBC in the absence of the leaking.

Note that $\Delta P$ is a no longer suitable quantity because $\Delta P=0$ for both nonchiral ($P_a=P_b=0$) and antichiral ($P_a=P_b\neq 0$) dynamics. We use the particle current to distinguish the antichiral dynamics from the chiral dynamics. In the operator representation, the influx of particles into the site $j$ of the leg-$\mu$ can be written as $\hat{I}_j^{\mu}=-i[\hat{\mu}_j^{\dagger}\hat{\mu}_j,\hat{H}]$. In the mean-field approximation, a straightforward calculation yields the following equations on the site-and-leg-dependent currents
\begin{gather}
	\begin{split}\label{influx a}
	{I}_j^{a}=&i[J_R^a e^{i\phi/2}({\psi_{j+1}^{a\dagger}}\psi_{j}^a-{\psi_{j}^{a}}^{\dagger}\psi_{j-1}^a) \\
&+J_L^a e^{-i\phi/2}({\psi_{j-1}^{a\dagger}}\psi_{j}^a-{\psi_{j}^{a}}^{\dagger}\psi_{j+1}^a)\\
&+K({\psi_{j}^{b}}^{\dagger}\psi_{j}^a-{\psi_{j}^{a}}^{\dagger}\psi_{j}^b)],
	\end{split}\\
 	\begin{split}\label{influx b}
 		{I}_j^{b}=&i[J_R^b e^{-i\phi/2}({\psi_{j+1}^{b\dagger}}\psi_{j}^b-{\psi_{j}^{b}}^{\dagger}\psi_{j-1}^b) \\
 &+J_L^b e^{i\phi/2}({\psi_{j-1}^{b\dagger}}{\psi_{j}^b-\psi_{j}^{b}}^{\dagger}\psi_{j+1}^b)\\
 		&+K({\psi_{j}^{a}}^{\dagger}\psi_{j}^b-{\psi_{j}^{b}}^{\dagger}\psi_{j}^a)],
 	\end{split}
\end{gather}
where the time-dependent wave functions are governed by the nonlinear Schr\"odinger equations in Eq. (\ref{nonlinear equation}). The first (second) and the last terms in Eqs. (\ref{influx a}) and (\ref{influx b}) denote the current toward the right (left) hand-side and the current between two legs, respectively. Hence, we can define the site-average current along leg-$\mu$ (at time $t$) as
\begin{equation}
\bar{C}_{\mu}=\frac{i}{L}\sum_{j}^{L-1} (J_L^{\mu} e^{-i\phi/2}{\psi_{j}^{\mu}}^{\dagger}\psi_{j+1}^{\mu} - J_R^{\mu} e^{i\phi/2}{\psi_{j+1}^{\mu\dagger}}\psi_{j}^{\mu}).
\end{equation}
For the chiral and antichiral dynamics, the currents along the two legs have opposite directions with $\bar{C}_a\bar{C}_b>0$ and the same direction with $\bar{C}_a\bar{C}_b<0$, respectively.

%%%%%%%%%%%%%%%%%%%%
\begin{figure}[t!]
	\centering
	\includegraphics[width=0.46\textwidth]{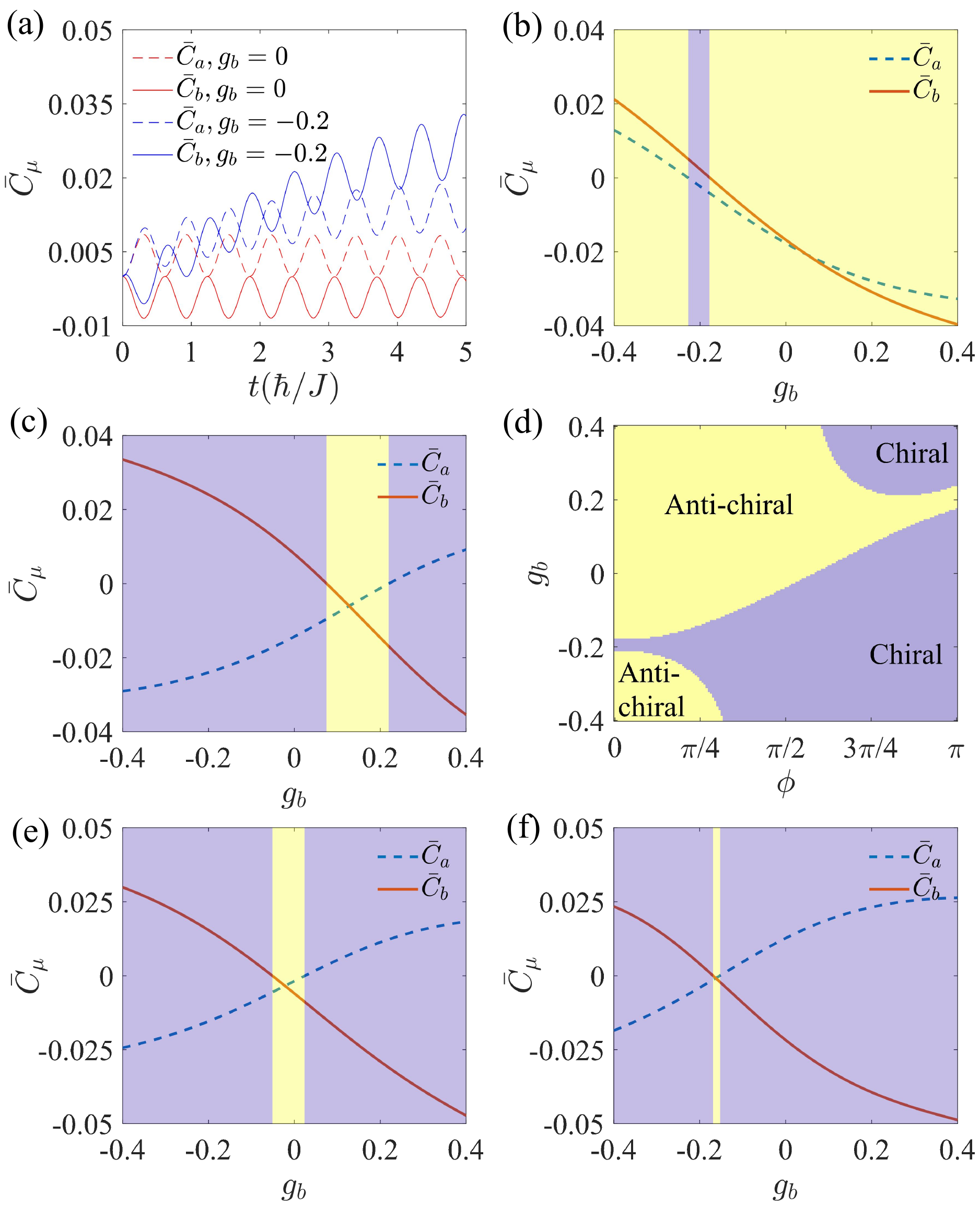}
	\caption{(Color online) (a) Time evolution of the site-average currents $\bar{C}_{\mu}$ along leg-$\mu$ for $g_b=0,-2$ and fixed $g_a=0$. $\bar{C}_{\mu}$ as functions of $g_b$ at $t=2$ for (b) $\phi=\pi/8$ and (c) $\phi=3\pi/4$. The yellow and purple shades denote the regions of $g_b$ for the antichiral and chiral dynamics, respectively. (d) Dynamical phase diagram in the $\phi$-$g_b$ plane. $\bar{C}_{\mu}$ versus $g_b$ at $t=2$ for (e) $U=2$ and (f) $U=4$. Other parameters are $U=0$ in (a-d), $g_a=0.2$ in (b-f), and $K=5$ in (a-f).}
	\label{fig3}
\end{figure}
%%%%%%%%%%%%%%%%%%%%

We first consider the noninteracting limit and depict typical results of $\bar{C}_{\mu}(t)$ in Fig.~\ref{fig3}(a) under the OBC. For the Hermitian case ($g_a=g_b=0$), one can find the chiral dynamics with $\bar{C}_a\bar{C}_b<0$ due to the opposite signs of the currents in two legs. For the nonreciprocal case with $g_a=0$ and $g_b=-0.2$, due to the NHSE (see the next section) on the leg-$b$, the currents on the two legs tend to the same direction and thus the antichiral dynamics with $\bar{C}_a\bar{C}_b>0$ is exhibited. Thus, the non-reciprocity can drive the system from the chiral into antichiral dynamical phases. We further study the chiral-antichiral transition by varying the magnetic flux $\phi$ and nonreciprocal hopping strength $g_b$ (with fixed $g_a=0.2$ without loss of generality). Since the antichiral and chiral dynamics are visible after a short time evolution (see Fig. \ref{fig3}(a) for instance), we choose $\bar{C}_{\mu}(t)$ at the evolution time $t=2$ (in the unit of $\hbar/J$) to identify different dynamics in Figs. \ref{fig3}(b-f). Two examples of $\bar{C}_{\mu}$ as functions of $g_b$ for the magnetic flux $\phi=\pi/8,3\pi/4$ are shown in Figs.~\ref{fig3}(b) and \ref{fig3}(c), respectively. By using the sign of $\bar{C}_a\bar{C}_b$, we also identify the chiral (purple shade) and antichiral (yellow shades) dynamics with respect to $g_b$. Note that one has $\bar{C}_a\bar{C}_b\approx0$ at the chiral-antichiral transition points. We numerically obtain the dynamical phase diagram in the $\phi$-$g_b$ plane, as shown in Fig.~\ref{fig3}(d). Notably, the chiral dynamics can be exhibited even for vanishing magnetic flux with $\phi=0$ when $g_b=-g_a=0.2$. This is due to the fact that the skin currents on two legs have opposite directions in this case. The chiral-antichiral transition is flux-tunable. As the flux $\phi$ is increased, the regions for the chiral and antichiral dynamics are enlarged and reduced, respectively. We further consider the nonlinear effect on the chiral-antichiral transition
under the nonreciprocal hopping. In Figs.~\ref{fig3}(e) and \ref{fig3}(f), we show the site-averaged currents $\bar{C}_{\mu}(t)$ at $t=2$ for $U=2$ and $U=4$, respectively. It is clear that when increasing the nonlinear interaction strength, the chiral-antichiral transition points shift and the antichiral region shrinks. As $U$ is further increased, the antichiral dynamics is suppressed and the chiral-antichiral transition disappears due to the nonlinear self-trapping effect.

\subsection{Non-Hermitian skin and trap-skin dynamics}

\begin{figure}[t!]
	\centering
	\includegraphics[width=0.46\textwidth]{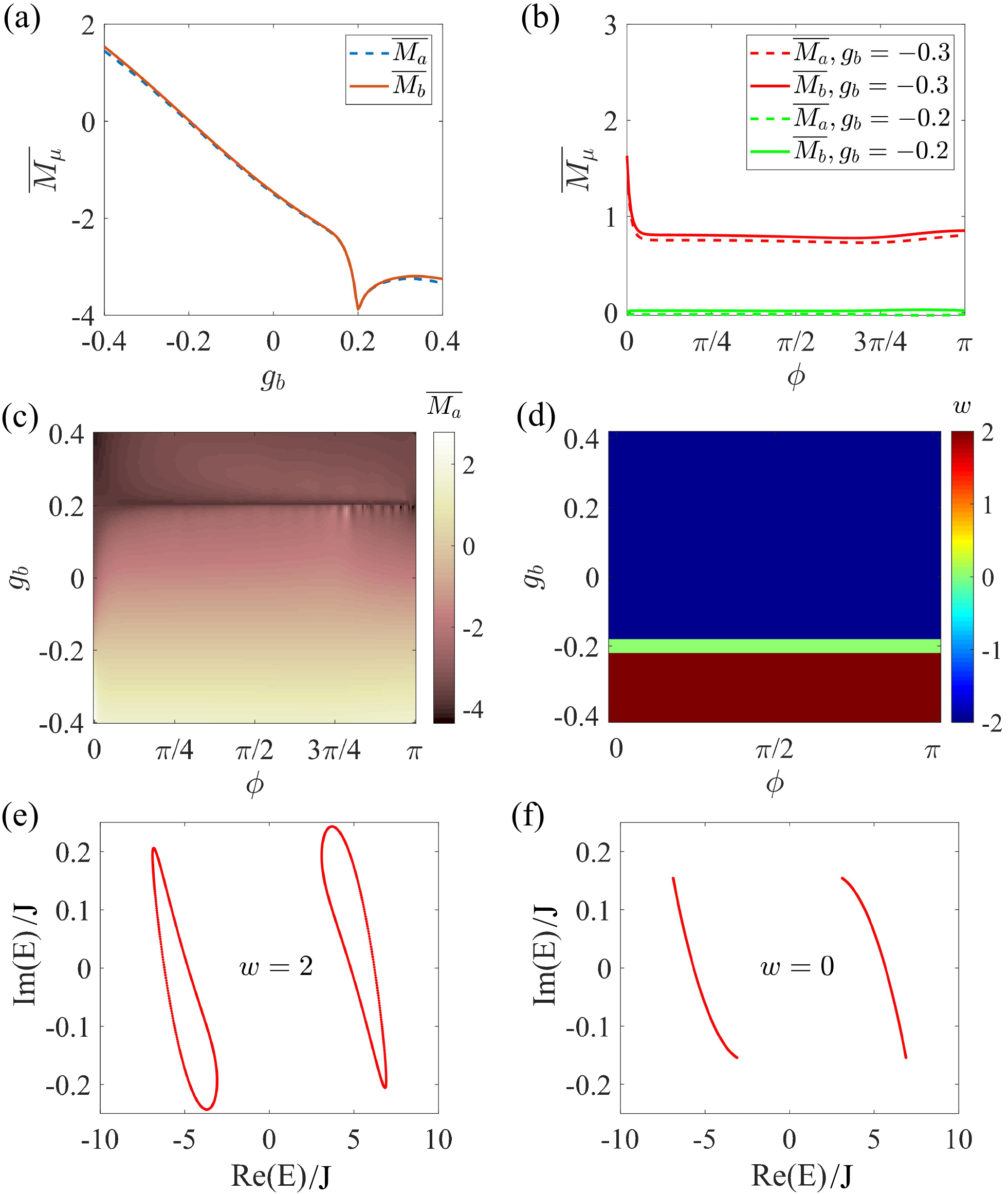}
	\caption{(Color online) (a) $\overline{M_{\mu}}$ versus $g_b$ for fixed $\phi=\pi/4$. (b) $\overline{M_{\mu}}$ versus $\phi$ for $g_b=-0.2,-0.3$. (c) $\overline{M_a}$ versus $\phi$ and $g_b$. (d) Spectral winding number $w$ versus $\phi$ and $g_b$. Complex energy spectra for (e) $\phi=\pi/4$ and $g_b=-0.3$ with $w=2$, and for (f) $\phi=\pi/4$ and $g_b=-0.2$ with $w=0$. Other parameters are $K=5$, $g_a=0.2$, and $U=0$.}
	\label{fig4}
\end{figure}

We proceed to study the NHSE under the magnetic flux and nonlinear interactions. The dynamic manifestation of the NHSE in our system is the skin dynamics, featuring the stacking of particle population at one end of the ladder. We define the time-averaged center-of-mass in leg-$\mu$ as
\begin{equation}\label{M1}
	\overline{M_{\mu}}=\frac{1}{T}\int_{0}^{T} \sum_{j}^{}j|\psi_j^{\mu}(t)|^2,
\end{equation}
where the evolution time $T=600$ is sufficiently long for skin modes accumulating onto the edges. We first consider the noninteracting case with $U=0$. The numerical result of $\overline{M_{a}}$ and $\overline{M_{b}}$ as functions of $g_b$ for fixed $g_a=0.2$ and $\phi=\pi/4$ is shown in Fig.~\ref{fig4}(a). We also plot $\overline{M_{a}}$ and $\overline{M_{b}}$ as functions of $\phi$ for $g_a=0.2$ and $g_b=-0.2,-0.3$ in Fig.~\ref{fig4}(b). We find that $\overline{M_{a}}\approx\overline{M_{b}}\neq0$ for different inter-leg tunneling strengths ($K=5$ in Fig. \ref{fig4} and others not shown here) as long as $g_b\neq -g_a$. Thus, the skin dynamics in two legs are almost the same, which leads to the antichiral dynamics discussed previously. The result of $\overline{M_{a}}$ ($\approx\overline{M_{b}}$) in the whole $\phi$-$g_b$ parameter plane in Fig.~\ref{fig4}(c) further demonstrates that the skin dynamics is always exhibited if $g_a\neq -g_b\neq0$ in the nonreciprocal ladder under the OBC and almost independent of the artificial magnetic flux. Note that $\overline{M_{\mu}}<0$ and $\overline{M_{\mu}}>0$ in Figs. \ref{fig4}(a-c) correspond to the left and right skin dynamics, respectively. When $g_b=-g_a$, the skin dynamics on two legs are canceled with $\overline{M_{\mu}}\approx0$, as the particle transports toward the left and right sides of the ladder are balanced.

The NHSE corresponds to the spectral topology in non-Hermitian systems~\cite{Gong2018}. The winding number with respect to a reference energy base in the complex energy plane for $U=0$ is defined as
\begin{equation}\label{winding}
	w=\sum_{n=1}^{2} \int_{0}^{2\pi} \frac{\partial_k}{2\pi} \arg \left[E_n(k)-E_B^{(n)}\right],
\end{equation}
where $E_n(k)$ are eigenenergies of the Bloch Hamiltonian $H(k)$ in Eq. (\ref{bloch H}) and $E_B^{(n)}$ are the energy bases for each band $n=1,2$. According to the non-Hermitian bulk-edge correspondence~\cite{Gong2018}, given a base energy $E_{B}$ that is not belonging to a single energy band under the PBC, a spectral winding number $w$ implies $|w|$ localized edge modes with energy $E=E_B$ under the OBC, and the sign of $w$ determines the left or right boundary for the skin mode. Thus, a nonzero winding number for a complex energy loop under the PBC indicates the presence of the NHSE under the OBC. Note that the winding number preserves for any base energy as long as it is within the closed loop, which is the only restriction on choosing the base energy. In numerical simulation, the energy bases $E_B^{(1)}$ and $E_B^{(2)}$ are independently chosen as (arbitrary) one of eigenenergies of each band under the OBC. As shown in Fig.~\ref{fig4}(d), one has $w=0$ only for $g_b=-g_a$ in the $\phi$-$g_b$ parameter plane, corresponding to $\overline{M_{\mu}}\approx0$ in Fig.~\ref{fig4}(a-c). Interestingly, one can see that the winding number is independent of the magnetic flux, which indicates the NHSE with skin modes is not suppressed by the magnetic flux in the ladder. This is in contrast to the magnetic suppression of NHSE in two-dimensional non-Hermitian systems~\cite{Lu2021,KShao2022}, where the number of bulk skin modes (the total winding number) is significantly reduced by the magnetic flux. Figures~\ref{fig4}(e) and \ref{fig4}(f) show the complex energy spectra and corresponding winding numbers for different values of $g_b$. There are two closed circles in the complex energy plane with total winding number $w=2$ when $g_a\neq -g_b$ [Fig.~\ref{fig4}(e)], corresponding to eigenstates being localized on the left side of the ladder under the OBC. For $g_a=-g_b$ in Fig.~\ref{fig4}(f), the total winding number $w=0$ for the two energy spectra confirms the absence of skin modes in this case.

%%%%%%%%%%%%%%%%%%%%
%%%%%%%%%%%%%%%%%%%%

%%%%%%%%%%%%%%%%%%%%
\begin{figure*}[t!]
	\centering
	\includegraphics[width=0.9\textwidth]{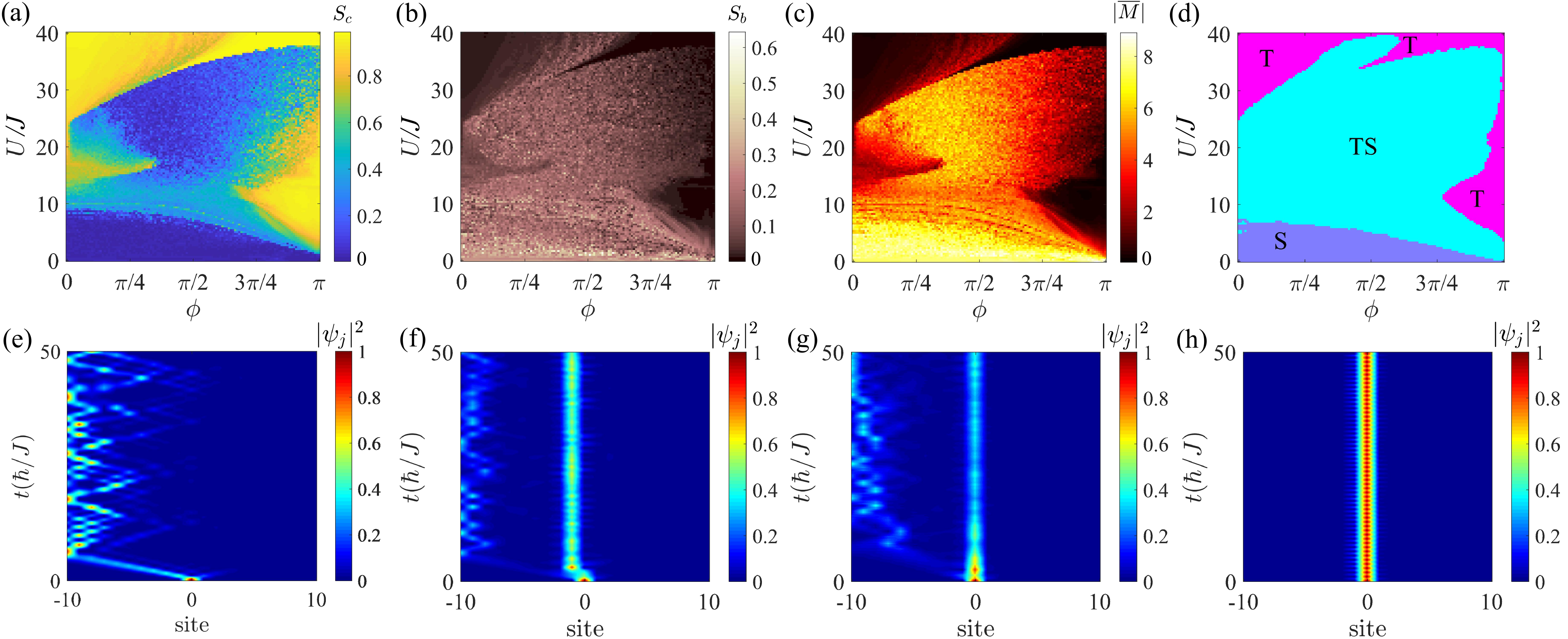}
	\caption{(a) $S_{c}$, (b) $S_{b}$, and (c) $|\overline{M}|$ as functions of $\phi$ and $U$. (d) Dynamical phase diagram in the $\phi$-$U$ parameter space. Here S denotes the skin dynamics, TS denotes the self-trapping dynamics, respectively. Time evolution of density distribution for (e) $U=0, \phi=0$; (f) $U=10, \phi=0$; (g) $U=10, \phi=\pi/2$; and (h) $U=10, \phi=\pi$. Other parameters are $K=5$ and $g_a=g_b=0.2$.}
	\label{fig5}
\end{figure*}
%%%%%%%%%%%%%%%%%%%%

We further consider the non-Hermitian skin dynamics under nonlinear interactions. For simplicity, we consider $g_a=g_b=0.2$ here, such that only leftward skin modes with finite $S_{j<0}$ are present. To reveal the interplay of the skin dynamics and the self-trapping dynamics, we use three quantities defined as
\begin{gather}
	S_{c}=S_0+S_{-1},\\
	S_{b}=S_{-10},\\
	|\overline{M}|=|\overline{M_a}|+|\overline{M_b}|.
\end{gather}
Here $S_{c}$ denotes the total density at the initial site $j=0$ and its left site, which can be used to characterize the self-trapping effect. $S_{b}$ denotes the left boundary mode at site $j=-10$ due to the NHSE. Under the interplay between the NHSE and the self-trapping effect, the wave packet may not be localized at the boundary. Thus, we further use the shift of the total center-of-mass $|\overline{M}|$ to reveal the complex dynamics. When the skin dynamics are dominant, $|\overline{M}|$ has relatively large values, while $|\overline{M}|<1$ if the self-trapping dynamics is dominated. As shown in Figs.~\ref{fig5}(a)-\ref{fig5}(c), we numerically calculate these three quantities in the $\phi$-$U$ plane. Combining these results, we obtain the dynamical phase diagram with three regions for the skin dynamics (denoted by S), the self-trapping dynamics (denoted by T), and the so-called trap-skin dynamics (denoted by TS) with the coexistence of trapped and skin modes, as shown in Fig.~\ref{fig5}(d). Note that here we use several cutoffs to estimate different dynamical phase regions: $S_{c}<0.05$ and $|\overline{M}|>6$ for the skin dynamics, $S_{b}<0.3$ and $|\overline{M}|<1$ for the self-trapping dynamics, and otherwise for the trap-skin dynamics. From Fig.~\ref{fig5}(d), one can find a large trap-skin parameter region, where the self-trapping and skin dynamics are both important. To see different dynamical behaviors more clearly, we plot the time evolution of the density distribution $|\psi_j|^2=|\psi_j^a|^2+|\psi_j^b|^2$ for various parameters $U$ and $\phi$ in Figs.~\ref{fig5}(e-f). There are three typical patterns distinguished from each other in the density evolution. The first pattern is the skin dynamics with skin modes near the left boundary shown in Fig.~\ref{fig5}(e). The second patterns are the trap-skin dynamics in Figs.~\ref{fig5}(f) and \ref{fig5}(g), where a partial density is dynamically localized near the boundary and the rest part is trapped near the initial site ($j=0$) or the nearby site ($j=-1$). The third pattern in Fig.~\ref{fig5}(h) shows the self-trapping dynamics around the initial position.

\section{\label{sec4} Discussion and Conclusion}

%%%%%%%%%%%%%%%%%%%%
\begin{figure}[t!]
	\centering
	\includegraphics[width=0.48\textwidth]{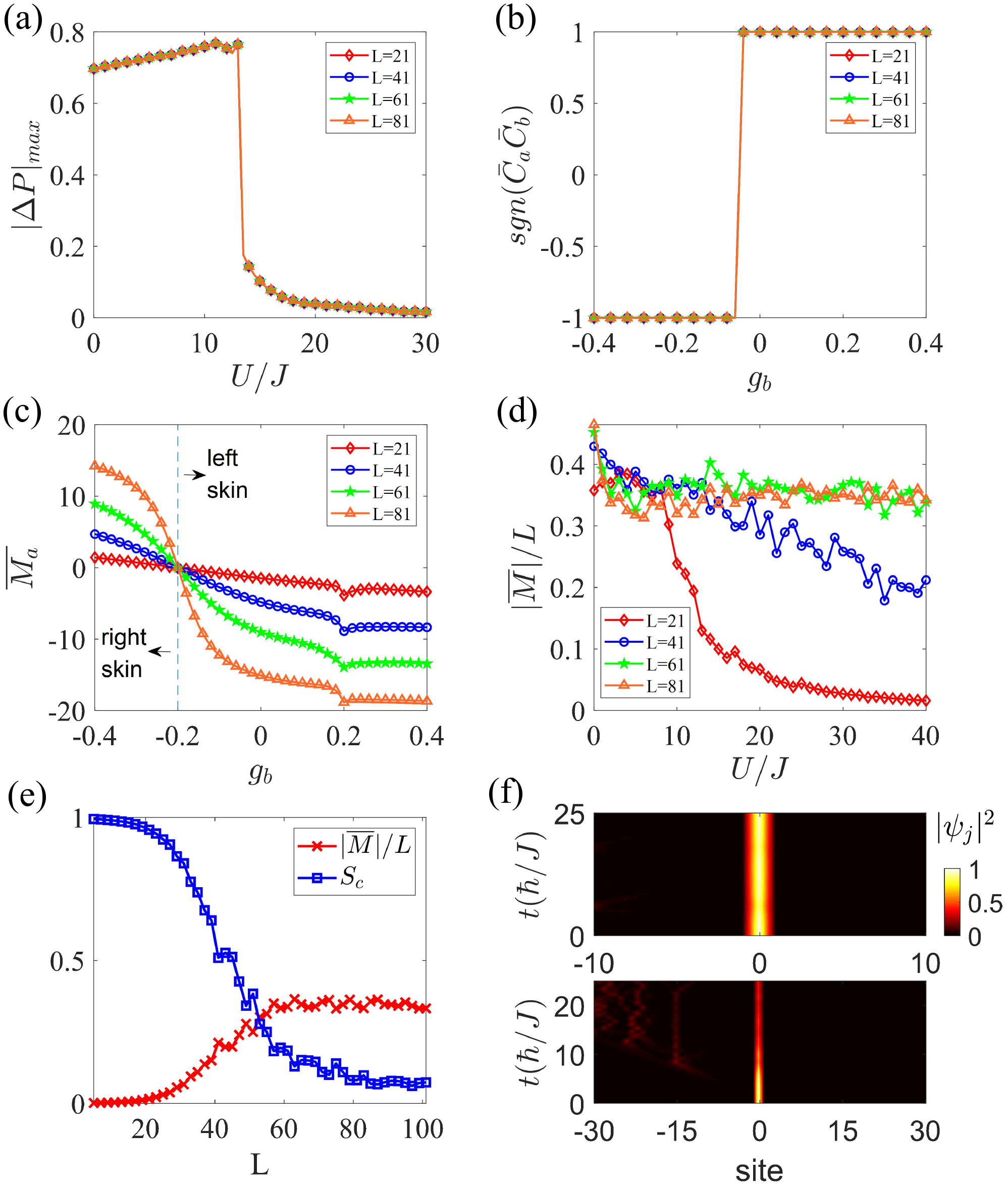}
	\caption{(a) $|\Delta P|_{\mathrm{max}}$ as a function of $U$ for different $L$ with $g_{a,b}=0$ and $K=0.8$. (b) $\text{sgn}(\bar{C}_a\bar{C}_b)$ and (c) $\overline{M_a}$ as functions of $g_b$ for different $L$ with $g_a=0.2$, $K=5$ and $U=0$. (d) $|\overline{M}|/L$ as a function of $U$ for different $L$. (e) $|\overline{M}|/L$ and $S_c$ as functions of $L$ with  $U=40$. (f) Time evolution of the density distribution for $L=21$ (top) and
$L=61$ (bottom) with $U=40$. Other parameters are $\phi=\pi/2$ in (a-c), $\phi=0$ in (d-f), and $K=5$ and $g_a=g_b=0.2$ in (b-f).}
	\label{fig6}
\end{figure}
%%%%%%%%%%%%%%%%%%%%

Before concluding, we briefly discuss the finite-size effect and show that our main results preserve for larger ladder lengths. Fig. \ref{fig6}(a) shows the maximum shearing $|\Delta P|_{\mathrm{max}}$ as a function of nonlinear interaction strength $U$ for ladder lengths $L=21,41,61,81$ in the Hermitian case. The values of $|\Delta P|_{\mathrm{max}}$ maintain for varying $L$. This indicates that the chiral dynamics for small (and moderate) $U$ and its breakdown due to the nonlinear self-trapping effect for large $U$ are unaffected for different ladder lengths. We also numerical calculate $\bar{C}_{\mu}$ and plot the sign of $\bar{C}_a\bar{C}_b$ as a function of the nonreciprocal strength $g_b$ in Fig. \ref{fig6}(b). One can see that both $\bar{C}_a\bar{C}_b<0$ for the chiral motion and $\bar{C}_a\bar{C}_b>0$ for the antichiral motion due to the NHSE preserve for different lengths, with the same chiral-antichiral transition points.
We further analyze the fine-site effect on the skin dynamics ($U=0$) and trap-skin dynamics ($U\neq0$). Figure \ref{fig6}(c) shows $\overline{M_a}$ as a function of $g_b$ for different $L$ in the noninteracting case. For a given $g_b$, as excepted, the center-of-mass of the skin modes in the leg-a ($\overline{M_a}$) near the left or right boundary becomes larger for longer ladders. However, the sign of $\overline{M_a}$ is unaffected for different $L$. This implies the left- and right-skin modes (with the wingding numbers $w=\pm2$) preserve in the noninteracting case, as noted in Fig. \ref{fig6}(c). To consider the interaction effect, we plot the rescaled center-of-mass $|\overline{M}|/L$ versus the nonlinear interaction strength $U$ with the flux $\phi=0$ in Fig. \ref{fig6}(d), which indicates the size-dependent skin dynamics when $U/J\gtrsim10$. In particular, the skin dynamics tends to be enhanced as $L$ is increased (up to $L=61,81$) for large $U$. The size-dependence of $|\overline{M}|/L$ and the self-trapping quantity $S_c$ for $U/J=40$ is shown in Fig. \ref{fig6}(e). As the ladder length is increased up to $L\lesssim21$, the nonlinear self-trapping dynamics with $S_c\sim1$ and $|\overline{M}|/L\sim0$ remain. With the increase of the ladder length up to $L\gtrsim61$, the values of $|\overline{M}|/L$ and $S_c$ become nearly stable with respect to $L$ and indicate the trap-skin dynamics. Typical time evolutions of the density distributions for the trapping dynamics with $L=21$ (top) and the trap-skin dynamics with $L=61$ (bottom) are shown in Fig. \ref{fig6}(f). Thus, in the strong interacting regime with the self-trapping effect, the skin modes can be exhibited when $L$ is sufficiently large. The parameter region for the trapping dynamics denoted by T (the trap-skin dynamics denoted by TS) in Fig. \ref{fig5}(d) slightly shrinks (enlarges) for larger $L$. Note that the size-dependent NHSE in non-Hermitian lattices without interactions was revealed in Ref.~\cite{li2020critical}. The finite-size effect on the interplay of the NHSE and (nonlinear and Hubbard) interactions requires further investigations.

In summary, we have explored the mean-field dynamics in a nonreciprocal bosonic two-leg ladder with artificial magnetic flux under the OBC. In the Hermitian limit, we have shown the breakdown of the chiral dynamics due to the nonlinear self-trapping effect with a crossover region. We have found that the nonreciprocity can drive the transition between chiral dynamics and antichiral dynamics. The antichiral motion is manifested as the non-Hermitian skin dynamics along the same direction on two legs without suppression by the magnetic flux, while the chiral-antichiral transition is flux-tunable. We have also revealed the trap-skin dynamics with coexistence of tapping and skin dynamics in the ladder. Several dynamical phase diagrams with respect to these different dynamics for the typical finite ladder have been obtained. These results showcase intriguing dynamical phenomena under the interplay among non-Hermiticity, nonlinearity, and artificial gauge fields. They could be experimentally tested with ultracold atoms since all the ingredients required for the model are already achieved therein.

\begin{acknowledgments}
This work is supported by the National Natural Science Foundation of China (Grants No. 12174126 and No. 12104166), the Guangdong Basic and Applied Basic Research Foundation (Grant No. 2024B1515020018), and the Science and Technology Program of Guangzhou (Grant No. 2024A04J3004).
\end{acknowledgments}

\bibliography{reference}

\end{document}